\documentclass[twocolumn,prl,aps,psfig,showpacs,preprintnumbersm,superscriptaddress]{revtex4}
\usepackage{graphicx}
\usepackage{epstopdf}
\usepackage{float}
\usepackage{color}
\usepackage{amsmath}

\begin{document}
\title{Correlation between Superconductivity and Magnetic Fluctuations in FeSe$_{1-x}$S$_x$ Revealed by $^{77}$Se NMR}

\author{P.~Wiecki}
\affiliation{Ames Laboratory, U.S. DOE and Department of Physics and Astronomy, Iowa State University, Ames, Iowa  50011  USA}
\author{K.~Rana}
\affiliation{Ames Laboratory, U.S. DOE and Department of Physics and Astronomy, Iowa State University, Ames, Iowa  50011  USA}
\author{A.~E.~B\"{o}hmer}
\affiliation{Ames Laboratory, U.S. DOE and Department of Physics and Astronomy, Iowa State University, Ames, Iowa  50011  USA}
\affiliation{Karlsruhe Institute of Technology, Institut f\"{u}r Festk\"{o}rperphysik, 76021 Karlsruhe, Germany}
\author{Y.~Lee}
\affiliation{Ames Laboratory, U.S. DOE and Department of Physics and Astronomy, Iowa State University, Ames, Iowa  50011  USA}
\author{S.~L.~Bud'ko}
\affiliation{Ames Laboratory, U.S. DOE and Department of Physics and Astronomy, Iowa State University, Ames, Iowa  50011  USA}
\author{P.~C.~Canfield}
\affiliation{Ames Laboratory, U.S. DOE and Department of Physics and Astronomy, Iowa State University, Ames, Iowa  50011  USA}
\author{Y.~Furukawa}
\affiliation{Ames Laboratory, U.S. DOE and Department of Physics and Astronomy, Iowa State University, Ames, Iowa  50011  USA}
\date{\today}

\begin{abstract}
   We present $^{77}$Se-NMR measurements on FeSe$_{1-x}$S$_x$ samples with sulfur content $x=0,9,15$ and $29\%$.  
    Twinned nematic domains are observed in the NMR spectrum for all samples except $x=29\%$.
    The NMR spin-lattice relaxation rate shows that magnetic fluctuations are initially enhanced between  $x=0\%$ and $x=9\%$, but are strongly suppressed for higher $x$ values. 
   The observed behavior of the  magnetic fluctuations parallels the superconducting transition temperature $T_c$ in these materials, providing strong evidence for the primary importance of magnetic fluctuations for superconductivity, despite the presence of nematic quantum criticality in this system.
\end{abstract}

\maketitle
 
 Critical fluctuations of an ordered phase found in the proximity to unconventional superconductivity have frequently been discussed as a source of superconducting
 pairing \cite{Scalapino2012,Curro2005,Hattori2012,Bertel2016}.
    In the iron-based superconductors \cite{Canfield2010,Johnston2010}, superconductivity (SC) is found in the vicinity of two types of long-range order: the stripe-type antiferromagnetic (AFM) order and the nematic order, which breaks in the in-plane rotational symmetry while preserving time reversal symmetry.
    While dynamical AFM fluctuations are well known to support SC, experimental and theoretical studies have suggested that nematic fluctuations 
may also be important for high-$T_c$ SC \cite{Kuo2016,Lederer2015,Lederer2017}. 

    In this context, FeSe has emerged as a key material since it undergoes a nematic phase transition from a tetragonal to an orthorhombic structure
    at $T_{\rm s}\approx90$ K and develops superconductivity below $T_{\rm c}\approx8.5$ K, but does not display static magnetic ordering \cite{Hsu2008,McQueen2009,Bohmer2018}.
    This suggests an opportunity to study the behavior of $T_c$ near a nematic quantum critical point (QCP) isolated from a magnetic QCP. 
    The nematic phase can be suppressed by pressure application, with $T_{\rm s}$ reaching 32 K at $p=1.5$ GPa. 
    However, an AFM ordered state emerges above $p=0.8$ GPa \cite{Terashima2015,Bendele2010} and merges with the nematic state above $p=1.7$ GPa \cite{Kothapalli2016}.
    Non-monotonic behavior of $T_{\rm c}$ is seen near the onset of the magnetic order \cite{Kaluarachchi2016}, but overall
    $T_{\rm c}$ is strongly enhanced up to 37 K at $p=6$ GPa \cite{Mizuguchi2008,Margadonna2009,Medvedev2009}. 
    While early nuclear magnetic resonance (NMR) measurements connected the enhancement of $T_{\rm c}$ to enhanced spin fluctuations under pressure \cite{Imai2009},
    the recently revealed complexity of the phase diagram raises new questions. 
    Notably, the role of nematic fluctuations in the superconductivity remains unclear. 

    The nematic phase can also be suppressed by S substitution in FeSe$_{1-x}$S$_x$  at ambient pressure, with the nematic phase disappearing around $x\approx17\%$. 
    Importantly, no long-range magnetic order can be observed at ambient pressure, which implies an isolated nematic QCP \cite{Hosoi2016}.
    $T_{\rm c}$ initially increases slightly to $T_c\approx10$ K at $x\approx10\%$ \cite{Abdel2015} from $T_{\rm c}\approx8.5$ K at $x = 0$,  but then decreases, reaching $T_c\approx5$ K by $x=29\%$. 
     The application of pressure induces magnetic order in S substituted samples \cite{Xiang2017,Matsuura2017}. 

     Recent results have highlighted the rich interplay between magnetic, nematic and superconducting orders in  the FeSe$_{1-x}$S$_x$ system. 
     Elastoresistivity measurements found that nematic fluctuations are divergently enhanced near the nematic QCP near $x\approx17\%$ \cite{Hosoi2016}.
     The full three-dimensional $T$-$p$-$x$ dependent phase diagram revealed strongly enhanced $T_{\rm c}$ in regions lacking both nematic and AFM long-range orders \cite{Matsuura2017}. 
     Furthermore, several studies have suggested that $T_{\rm c}$ does not appear to correlate with nematicity in FeSe$_{1-x}$S$_x$ \cite{Hosoi2016,Matsuura2017,Watson2015,Coldea2016}. 
      On the other hand, no direct measurements of the concentration dependence of magnetic fluctuations have been reported yet.

     Since magnetic fluctuations are considered to be one of the key ingredients  for the appearance of SC in iron pnictides, it is crucial to reveal how magnetic fluctuations  vary with S substitution  in FeSe$_{1-x}$S$_x$.
    NMR is an ideal tool for the microscopic study of low-energy magnetic fluctuations in correlated electron systems.
    Here, we carried out  $^{77}$Se NMR measurements to investigate static and dynamic magnetic properties of FeSe$_{1-x}$S$_x$.
    Our NMR data clearly show that stripe-type AFM fluctuations are initially slightly enhanced by S doping up to  $x\approx10\%$ from $x$ = 0 but are 
strongly suppressed thereafter, particularly beyond the nematic dome above $x\approx17\%$. 
    This behavior shows a strong correlation with $T_{\rm c}$, providing clear evidence for the primary importance of AFM fluctuations 
    over critical nematic fluctuations for SC in the FeSe$_{1-x}$S$_x$ system.
 

   $^{77}$Se NMR measurements have been carried out under a fixed magnetic external field of $H=7.4089$ T applied either along the $c$ axis or in the $ab$ plane ([110] tetragonal direction). 
    The crystals were grown using chemical vapor transport as outlined in Ref. \cite{Bohmer2016,SM}.  
     The four different S-content crystals used in this study are $x =0$ ($T_{\rm s}$ = 90 K, $T_{\rm c}$ = 8.5 K), $x =0.09$ ($T_{\rm s}$ = 68 K, $T_{\rm c}$ = 10 K), 
$x =0.15$ ($T_{\rm s}$ = 45 K, $T_{\rm c}$ = 8 K),   and $x =0.25$ ($T_{\rm c}$ = 5 K).
     Further experimental details are described in the Supplemental Material (SM) \cite{SM}.

\begin{figure}[tb]
\centering
\includegraphics[width=\columnwidth]{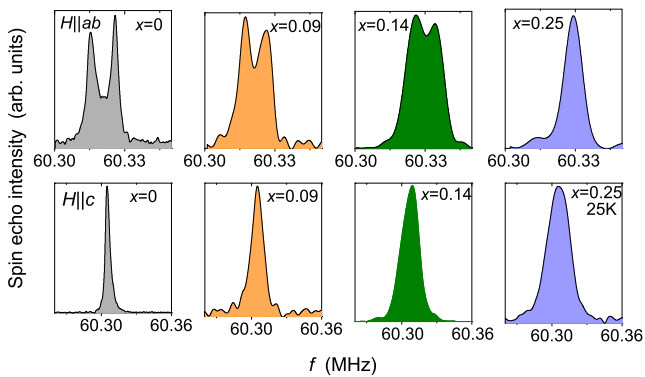}
\caption{Representative NMR spectra with $H||ab$ (upper panels) and $H||c$ (lower panels) at $T=20$ K (unless otherwise specified) for indicated S concentrations $x$. 
}
\label{fig:spectra}
\end{figure}

\begin{figure}[b]
\centering
\includegraphics[width=\columnwidth]{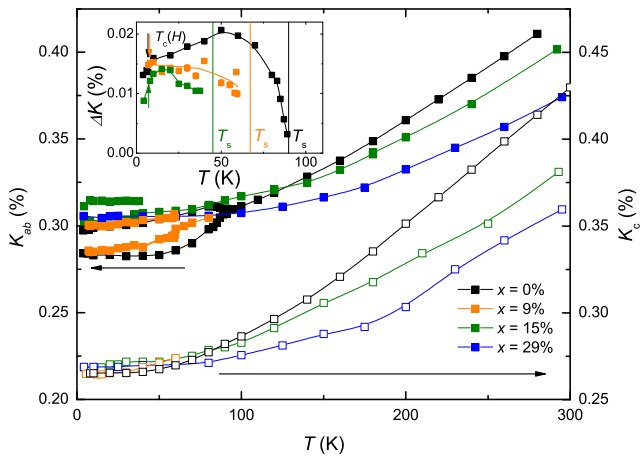}
\caption{$T$ dependence of the NMR shift $K$ for indicated S concentrations $x$ for external fields $H||ab$ (filled symbols) 
and $H||c$ (open symbols). 
Inset: Splitting $\Delta K$ of the $H||ab$ NMR spectrum due to twinned nematic domains. 
Vertical lines represent $T_{\rm s}$ determined by resistivity measurements \cite{SM}.
Arrows in inset represent $T_c(H)$ determined by {\it in situ} ac-susceptibility \cite{SM}. 
}
\label{fig:K}
\label{fig:deltaK}
\end{figure}

     In pure FeSe, the single peak observed in the $H||ab$ NMR spectrum at high $T$ splits into two peaks below $T_{\rm s}$ due to nematic order, where the two peaks arise from the presence of twinned nematic domains \cite{Baek2015,Bohmer2015,Wiecki2017}.  
  Representative NMR spectra at $T=20$ K for both field directions are shown in Fig. \ref{fig:spectra}. 
    Splittings of the $H||ab$ spectra below $T_{\rm s}$ are also observed in  FeSe$_{1-x}$S$_x$ except for $x=29\%$ where only a single peak is observed down to the lowest temperature, consistent with the lack of nematic order seen by resistivity \cite{SM}.

    The $T$ dependence of the NMR shift $K$ for all samples and both $H$  directions is shown in Fig. \ref{fig:K}. 
     As in pure FeSe, all $K$ values increase monotonically with increasing $T$. 
   $K_{ab}$ is greater than $K_c$ for all samples with almost no $x$ dependence  at low $T$.    
    On the other hand, the high temperature value of $K$ shows a large concentration dependence, where $K$ decreases with  increasing  $x$.

     The inset of Fig. \ref{fig:deltaK} shows the $T$ and $x$ dependence of the $H||ab$ spectral splitting $\Delta K$ (the difference of the Knight shifts of the two peaks), which is a measure of the local  nematic order parameter \cite{Baek2015}.
    For the pure sample,  $\Delta K$ increases sharply below $T_{\rm s}$  and shows a broad maximum near $\sim50$ K, as reported previously \cite{Baek2015,Wang2016,Wiecki2017}.
    In contrast to pure FeSe,  $\Delta K$ for $x$ = 9\% and $x$ = 15\% does not exhibit this maximum. 
While the $\Delta K$ of the $x$ = 0\% and $x$ = 9\% samples show no clear kinks at $T_c$, the $x$ = 15\% sample shows a noticeable drop in the SC state. 
   In the S-doped samples, we could not resolve the splitting all the way up to the bulk $T_{\rm s}$ identified by resistivity measurements \cite{SM},
     likely due to the broadening of the two individual lines (see Fig. \ref{fig:spectra}) by microscopic disorder from dopants and/or small  variations in the local S composition. 
     Due to the broad spectra relative to pure FeSe, no clear evidence for the local nematicity above $T_{\rm s}$, observed in pure FeSe from FWHM measurements \cite{Wiecki2017,Wang2017}, could be found. 

\begin{figure}[t]
\centering
\includegraphics[width=0.9\columnwidth]{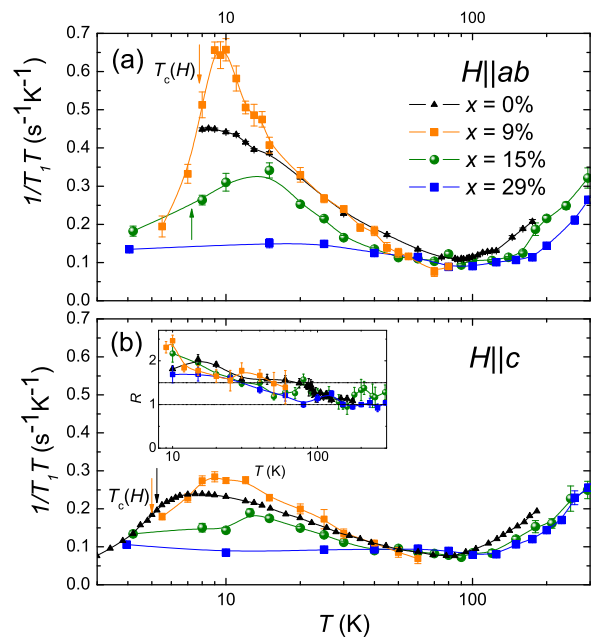}
\caption{$T$ dependence of NMR spin-lattice relaxation rate $1/T_1T$ for $H||ab$ (upper panel) and $H||c$ (lower panel)
for indicated S concentrations $x$. 
Arrows denote observed $T_c(H)$ as determined from {\it in situ} ac-susceptibility \cite{SM} (not shown for $x=0\%$ $H\|ab$). 
For S-doped samples, missing arrows indicate $T_c(H)<4.0$ K. 
Inset: The $T$ dependence of the anisotropy ratio $R=T_{1,c}/T_{1,ab}$ above $T_{\rm c}$ (see text). 
Data for $x=0\%$ ($ab$ plane average $1/T_1T$ and ratio $R$ at $H=9$ T) are from Ref. \onlinecite{Bohmer2015}. 
Data for $x=0\%$ ($H||c$ at $H=7$ T) are from Ref. \onlinecite{Shi2018}.
}
\label{fig:T1T}
\end{figure}

    We now discuss the behavior of the low-energy magnetic fluctuations based the NMR spin-lattice relaxation rate ($1/T_1$) data. 
    $1/T_1T$ for all samples and both $H$ directions are shown in Fig. \ref{fig:T1T} \cite{T1_ab}.
     In general, 1/$T_1T$ is related to the dynamical magnetic susceptibility as $1/T_1T\sim\gamma^{2}_{N}k_{\rm B}\sum_{\mathbf{q}}|A(\mathbf{q})|^2\chi^{\prime\prime}(\mathbf{q}, \omega_N)/\omega_N$, where $A(\mathbf{q})$ is the wave-vector $\mathbf{q}$ dependent form factor and $\chi^{\prime\prime}(\mathbf{q}, \omega_N)$ is the imaginary part of the dynamic susceptibility at the Larmor frequency $\omega_N$ \cite{Smerald2011}. 
    Above $\sim100$ K, $1/T_1T$ shows a similar $T$ dependence as the NMR shift $K(T)$ which measures the uniform susceptibility $\chi(\mathbf{q}=0)$. 
    In contrast, below $\sim100$ K a strong upturn of $1/T_1T$ is observed which is not seen in $K(T)$. 
   The enhancement of $1/T_1T$ at low $T$ is therefore attributed to the growth of AFM spin fluctuations with $\mathbf{q}\neq0$. 
    The AFM fluctuations appear below $\sim100$ K for all samples, but the enhancement of the AFM fluctuations shows a strong $x$ dependence.


   In order to characterize the AFM fluctuations, we plotted the ratio of 1/$T_1$ for the two field directions, $R\equiv(1/T_1T)_{ab}/(1/T_1T)_{c}=T_{1,c}/T_{1,ab}$. 
  According to previous NMR studies performed on Fe pnictides and related materials \cite{Kitagawa2009,Kitagawa2010,Hirano2012, Furukawa2014,Pandey2013,Ding2016}, $R$ depends on the wavevector of the spin correlations. Assuming isotropic spin correlations, one expects $R=1.5$ for stripe-type, $R=0.5$   for N\'{e}el-type. 
    As plotted in the inset of Fig.\ \ref{fig:T1T}(b), $R$ $\approx$ 1 at high $T$ and increases to $R>1.5$ starting below $\sim100$ K. 
    The value of $R$ observed here at low $T$ is consistent with stripe-type spin correlations. 
      The $T$ dependence of $R$ is independent of doping $x$ within experimental error, indicating no change in the character of magnetic fluctuations with doping. 

\begin{figure}[tb]
\centering
\includegraphics[width=\columnwidth]{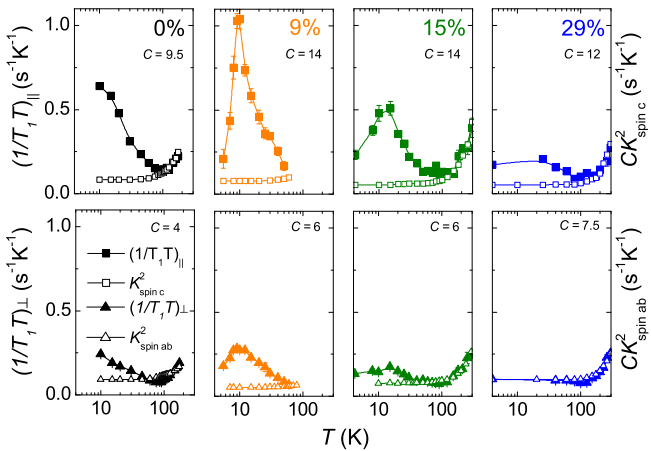}
\caption{
Comparison of $1/T_1T$ (left axes, filled symbols) with $CK_{{\rm spin}}^2$ (right axes, open symbols) for indicated doping levels. The upper panels compare
$1/T_{1,\|}T$ to $CK_{{\rm spin},c}^2$, while the lower panels compare  $1/T_{1,\perp}T$ to $CK_{{\rm spin},ab}^2$ (see text).
The empirical value of $C$ (in units of $10^4$s$^{-1}$K$^{-1}$) for each panel is indicated (see \cite{SM}).
}
\label{fig:T1TKsqr}
\end{figure}

\begin{figure*}[t]
\centering
\includegraphics[width=\textwidth]{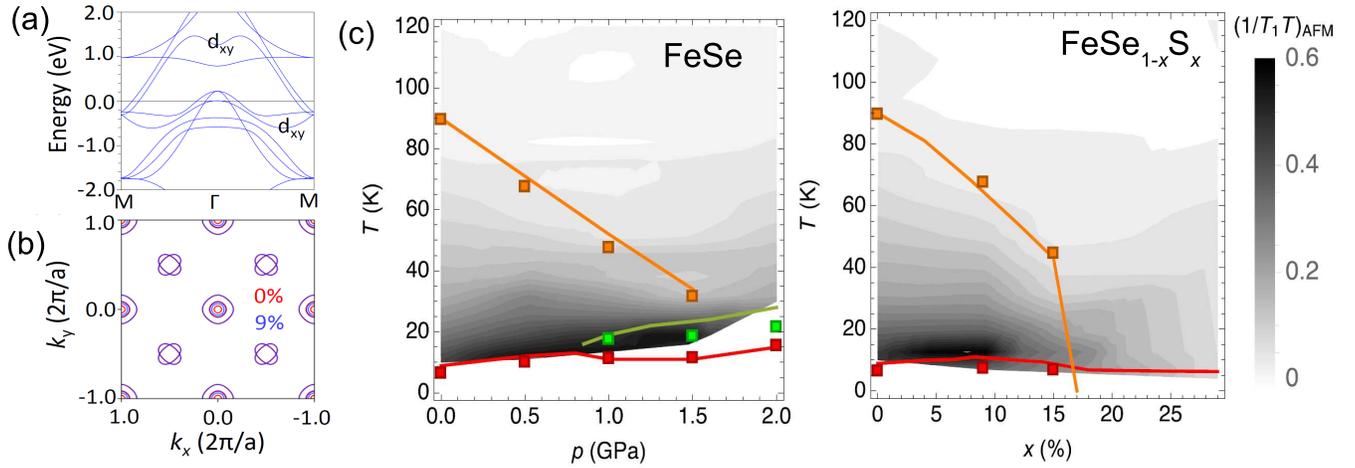}
\caption{
(a) Band dispersion of pure FeSe in the tetragonal phase, with bands of $d_{xy}$ orbital character indicated. 
(b) Cross-sections of the Fermi surface in the tetragonal phase at $k_{\rm z}$ = 0 for $x=0\%$ (red) and $x=9\%$ (blue).
(c) Comparison of AFM fluctuations in pure FeSe under pressure \cite{Wiecki2017} (left panel) and the FeSe$_{1-x}$S$_x$ system (right panel). 
Here, the AFM contribution to $1/T_1T$ is defined by $(1/T_1T)_{\rm AFM}\equiv(1/T_1T)-(1/T_1T)_{\mathbf{q}=0}$ using $H\|ab$ data \cite{SM}.
Solid lines show  $T_{\rm s}$ (orange), $T_{\rm N}$ (green) and $T_{\rm c}$ (red) determined from resistivity \cite{Kaluarachchi2016,Terashima2015,Reiss2017,Xiang2017}. 
Data points show $T_{\rm s}$, $T_{\rm N}$ and $T_{\rm c}(H)$ from NMR (this work and \cite{Wiecki2017}).
}
\label{fig:contour}
\end{figure*}

    To discuss magnetic fluctuations in more detail, it is convenient to isolate the component-resolved hyperfine field (HF) fluctuations from the measured $1/T_1$ data. 
     $1/T_1$ probes the $\mathbf{q}$ sum of fluctuations of HF at $\omega_N$ perpendicular
to the applied field according to   
$(1/T_1)_{H||i}=\gamma_{\rm N}^2\sum_\mathbf{q}\left[|H^{\rm hf}_j(\mathbf{q},\omega_{\rm N})|^2+|H^{\rm hf}_k(\mathbf{q},\omega_{\rm N})|^2\right]$,
where $(i,j,k)$ are mutually orthogonal directions and $|H^{\rm hf}_j(\mathbf{q},\omega)|^2$ represents the 
$\mathbf{q}$-dependent power spectral density of the $j$-th component of HF at the nuclear site. 
Therefore, we define the quantities 
$1/T_{1,\perp}\equiv(1/T_1)_{H||c}=2\gamma_{\rm N}^2\sum_\mathbf{q}|H^{\rm hf}_{ab}(\mathbf{q},\omega_{\rm N})|^2 $
and $1/T_{1,\|}\equiv2(1/T_1)_{H||ab}-(1/T_1)_{H||c}=2\gamma_N^2\sum_\mathbf{q}|H^{\rm hf}_{c}(\mathbf{q},\omega_{\rm N})|^2$ \cite{Wiecki2015}.
      Note that, for simplicity, we have neglected any $ab$-plane anisotropy due to nematicity ($H^{\rm hf}_{a}=H^{\rm hf}_{b}\equiv H^{\rm hf}_{ab} $).
      Thus defined, $1/T_{1,\perp}$ ($1/T_{1,\|}$) directly measures the $ab$ ($c$) component of HF fluctuations 
$\sum_\mathbf{q}|H^{\rm hf}_{ab}(\mathbf{q},\omega_{\rm N})|^2$ ($\sum_\mathbf{q}|H^{\rm hf}_{c}(\mathbf{q},\omega_{\rm N})|^2$). 

     In Fermi liquid systems, one expects that $1/T_1T\propto K_{\rm spin}^2$. 
    Here $K_{\rm spin}=K-K_0$, where $K_0$ is the $T$-independent chemical shift. 
    $K_{\rm spin}$ probes the uniform $\mathbf{q}=0$ susceptibility according to  $K_{{\rm spin},i}=A_{ii}\chi_{ii}(\mathbf{0})$, where $A_{ii}$ is the  hyperfine coupling constant. 
         Therefore, to examine the contribution of $\mathbf{q}\neq0$ correlations one can compare $1/T_1T$ to $K_{\rm spin}^2$. 
         The quantities $1/T_{1,\|}T$  and $1/T_{1,\perp}T$ should be compared to $K_{{\rm spin},c}^2$ and $K_{{\rm spin},ab}^2$, respectively \cite{Wiecki2015}.
     The experimentally observed $1/T_1T$ can then be decomposed into $\mathbf{q}=0$ and AFM ($\mathbf{q}\neq0$) components
as  $1/T_1T=(1/T_1T)_{\rm AFM}+(1/T_1T)_{\mathbf{q}=0}$. We have $(1/T_1T)_{\mathbf{q}=0}=CK_{\rm spin}^2$, where $C$ is a proportionality constant 
determined empirically from the high $T$ data \cite{SM}. 

    In Fig. \ref{fig:T1TKsqr}, we compare the angle-resolved pairs of $1/T_1T$ and $CK_{\rm spin}^2$.
     Above $\sim100$ K, it is clear that $1/T_1T\approx CK_{\rm spin}^2$, indicating that the $T_1$ relaxation  is being driven primarily by the $\mathbf{q}=0$ component.    
    In contrast, the difference between $1/T_1T$ and $CK_{\rm spin}^2$ can be clearly seen below $\sim100$ K and is attributed to the contribution from the stripe-type AFM fluctuations, $(1/T_1T)_{\rm AFM}$.
    Relative to pure FeSe, spin fluctuations are enhanced at $x=9\%$, slightly suppressed at $x=15\%$ and strongly suppressed for $x=29\%$.
     The $x$ dependence of the magnetic-fluctuation enhancement closely parallels the $x$ dependence of $T_{\rm c}$, which shows a slight enhancement between $x=0\%$ and $x=9\%$ and is suppressed at higher doping levels. 
    The suppression of magnetic fluctuations for $x\geq15\%$ is consistent with ARPES data \cite{Reiss2017}. 

    In all cases, we find that $1/T_{1,\|}T>1/T_{1,\perp}T$ at low $T$, indicating that  $\sum_\mathbf{q}|H^{\rm hf}_c(\mathbf{q},\omega_{\rm N})|^2$ 
    is greater than $\sum_\mathbf{q}|H^{\rm hf}_{ab}(\mathbf{q},\omega_{\rm N})|^2$. 
   The hyperfine field at the Se nuclear site is determined from the magnetic moments on the Fe sites  by the hyperfine coupling tensor. 
    Since the stripe-type AFM fluctuations produce the HF fluctuations at the  Se site though off-diagonal components of the hyperfine coupling tensor \cite{Kitagawa2008,Johnston2010}, the fact that $|H^{\rm hf}_{c}|^2$ is greater than  $|H^{\rm hf}_{ab}|^2$ shows that the $ab$-plane polarized stripe-type AFM fluctuations are more developed than the 
    corresponding $c$-axis polarized fluctuations, similar to the BaFe$_2$As$_2$-based superconductors \cite{Wiecki2015}. 

Within an itinerant picture, the change in the AFM spin correlations with doping would be associated with a change in the nesting condition due to 
modification of the Fermi surface with S substitution.
    To understand the band structure of FeSe$_{1-x}$S$_x$, we performed electronic structure calculations \cite{Yongbin}  using the full-potential linearized augmented plane wave  method \cite{Blaha2001} with a generalized gradient approximation \cite{Perdew1996}.
    Here we calculate the band structure for the tetragonal phases in  FeSe$_{1-x}$S$_x$ using an FeSe unit cell,
    adopting chemical pressure effects on the $a$ and $c$ lattice parameters. 
      The calculated band dispersion is shown in Fig. \ref{fig:contour}(a), which is in good agreement with the previous report \cite{Watson2015}.
   The calculated Fermi surface has three hole pockets around the $\Gamma$ point and two electron pockets at the $M$ point along the [110] direction (Fig. \ref{fig:contour}(b)). 
    We find that  the size of the smallest of the three hole pockets, originating from the $d_{xy}$ orbital, is increased by S doping. In contrast, the other pockets, 
    originating from $d_{yz}$ and $d_{zx}$ orbitals, do not change.
    These results continue to hold for a $1\%$ reduction of the chalcogen height, which also occurs by S doping \cite{Matsuura2017}.
    Thus the $d_{xy}$ orbital can be considered to play an important role in AFM spin correlations and also in the appearance of SC in FeSe$_{1-x}$S$_x$.

    Finally let us comment on the temperature dependence of $1/T_1T$ observed in $x=9\%$ and $x=15\%$ (see Fig. \ref{fig:T1T}). 
    For $x=0\%$, the maximum of $1/T_1T$ has been reported to occur close to $T_{\rm c}$ \cite{Baek2015,Wang2016}. 
    However, for $x=9\%$ and $x=15\%$, we find that the maximum of  $1/T_1T$ instead occurs well above $T_{\rm c}(H)$ as determined by our {\it in situ} ac-susceptibility measurements \cite{SM}. 
     At $x=9\%$, we find $T_{\rm c}(H||ab)=7.8$ K and $T_{\rm c}(H||c)=5.0$ K, while $1/T_1T$ peaks at $\sim9$ K for both $H$ directions. 
     At $x=15\%$, we find $T_{\rm c}(H||ab)=7.25$ K and $T_{\rm c}(H||c)\leq4.0$ K. 
     However, for both $H$ directions, $1/T_1T$ peaks at $\sim12-15$ K.
      These results imply a suppression of magnetic fluctuations just above $T_{\rm c}$ in the S-doped samples. 
      The effect is more apparent for $H||c$ data. Furthermore, the $T$ difference between $T_c$ and the peak of $1/T_1T$ appears to increase
      with doping.
     It is interesting to point out that similar behavior has been observed in pure FeSe and discussed in terms of a 
     possible superconducting fluctuation effect \cite{Kasahara2016,Shi2018}. 
     Detailed field-dependent measurements on the S-doped samples will be needed to confirm this scenario. 


     Our main results are summarized in the phase diagram of Fig. \ref{fig:contour}(c), which shows a contour plot of the AFM contribution to $1/T_1T$ as a function of $x$ and $T$. 
   For comparison, a similar plot for pure FeSe under pressure is also shown. 
   In both cases, the bulk nematic order is suppressed. 
   In pure FeSe, the magnetic fluctuations are roughly independent of pressure or slightly enhanced.
   In contrast, magnetic fluctuations are ultimately strongly suppressed by S doping, after an initial slight enhancement for $x\approx9\%$. 
   Magnetic fluctuations are strongly correlated with $T_{\rm c}$  in the FeSe$_{1-x}$S$_x$ system. 
   In contrast, nematic fluctuations are most strongly enhanced near the nematic critical quantum point at $x\approx17\%$ \cite{Hosoi2016} and show no correlation with $T_{\rm c}$. 
   These NMR results demonstrate the primary importance of magnetic fluctuations to superconductivity in the FeSe system, and help to disentangle the roles of magnetic and nematic fluctuations in iron-based superconductors in general. 
  
      The research was supported by the U.S. Department of Energy, Office of Basic Energy Sciences, Division of Materials Sciences and Engineering. Ames Laboratory is operated for the U.S. Department of Energy by Iowa State University under Contract No.~DE-AC02-07CH11358.

\newpage

\clearpage

\section{Supplementary Material}

\setcounter{figure}{0}

\begin{figure}[b]
	\centering
	\includegraphics[width=0.9\columnwidth]{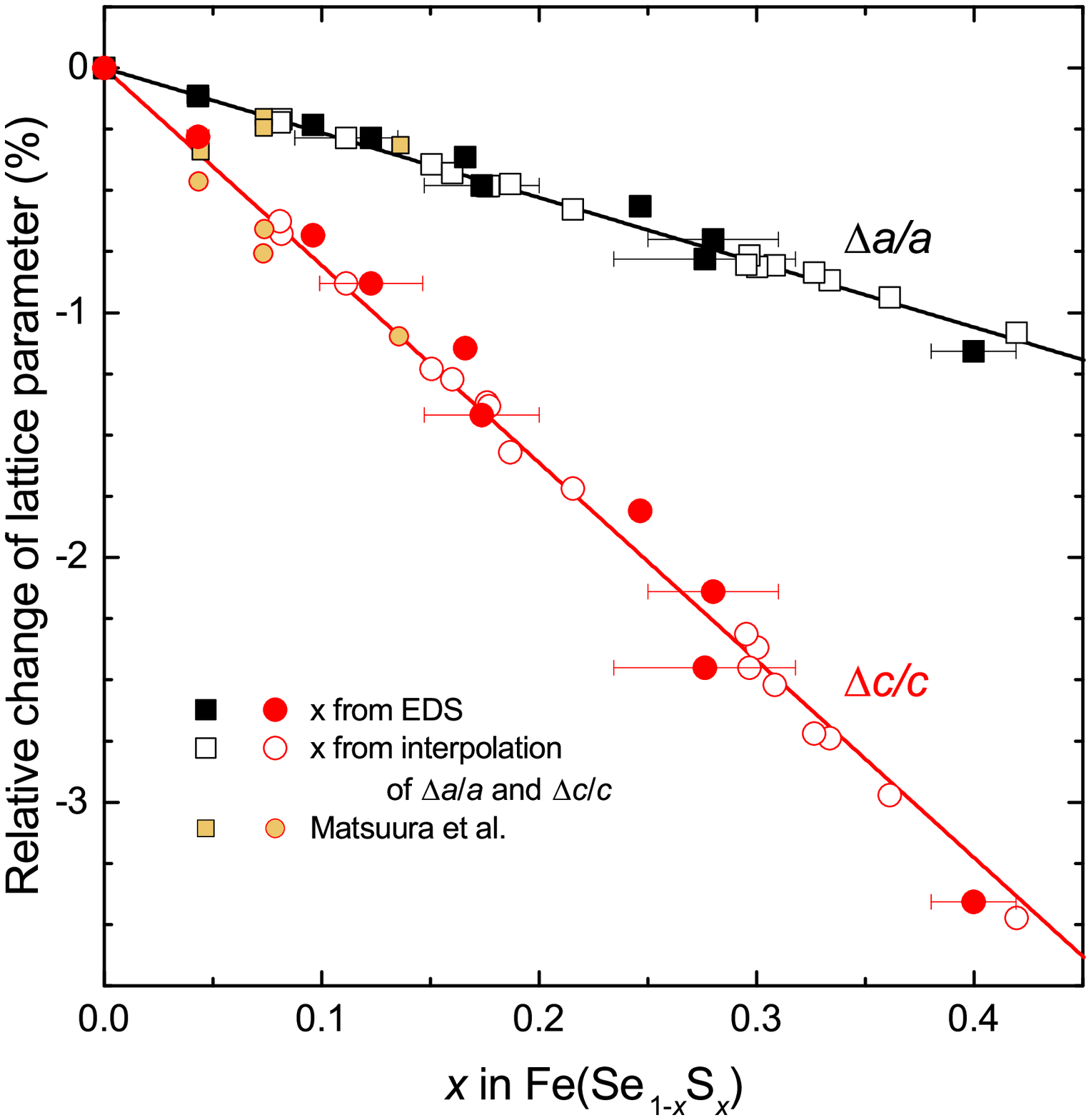}
	\caption{Change of $a$- and $c$-axis lattice parameters of FeSe$_{1-x}$S$_x$ with sulfur content $x$ from room-temperature powder diffraction data. For each batch shown as a full symbol, $x$ was determined by energy-dispersive x-ray spectroscopy (EDS). Open symbols have been placed by interpolating both $a$ and $c$ lattice parameters. Literature data from Ref. \onlinecite{Matsuura2017} has been added for comparison.} 
	\label{fig:latticeparameters}
\end{figure}

\begin{figure}[b]
	\centering
	\includegraphics[width=\columnwidth]{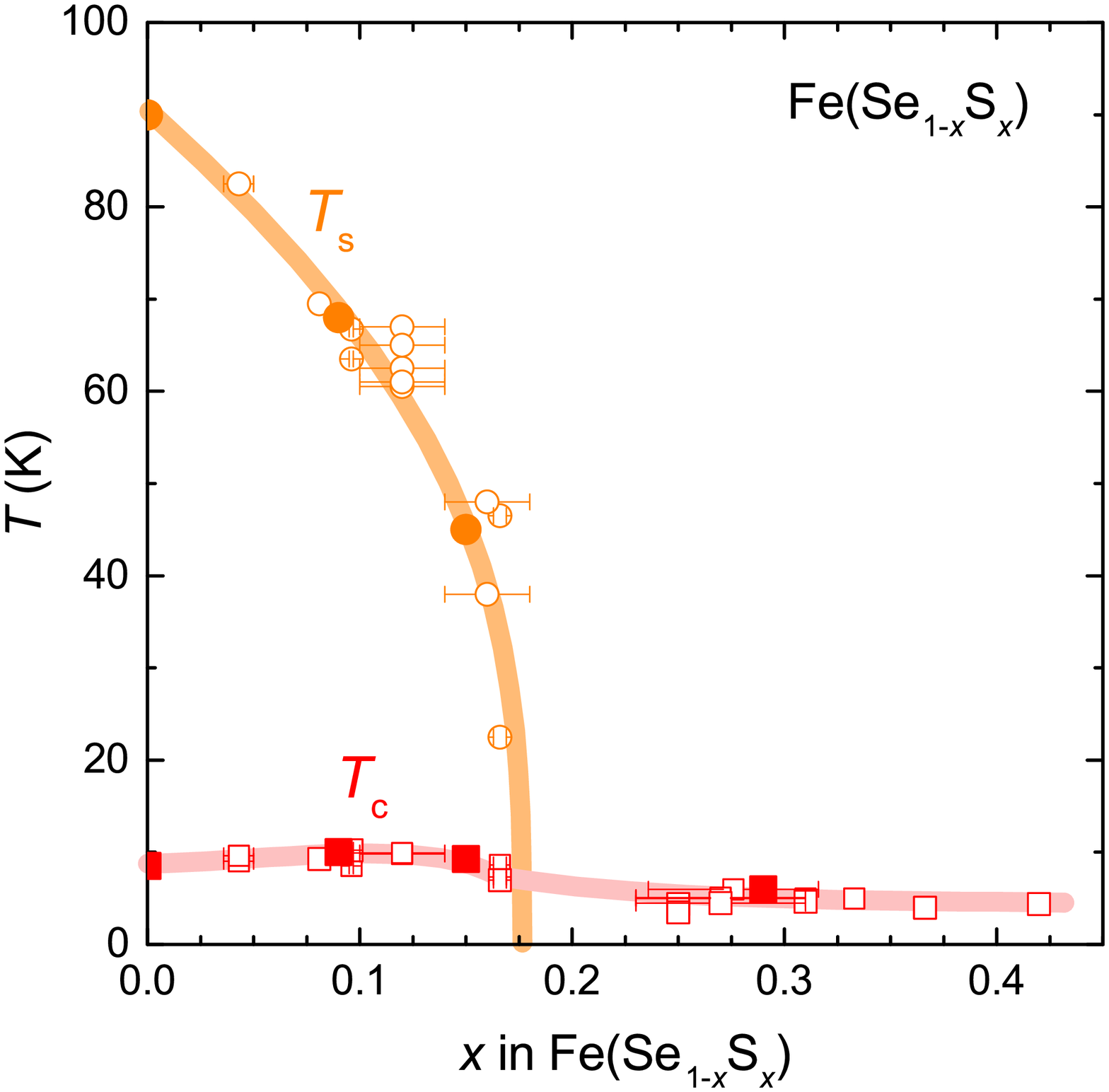}
	\caption{Phase diagram of FeSe$_{1-x}$S$_x$. The samples investigated by NMR in the main paper are highlighted by full symbols.}
	\label{fig:phasediag}
\end{figure}

\begin{figure}[tb]
	\centering
	\includegraphics[width=\columnwidth]{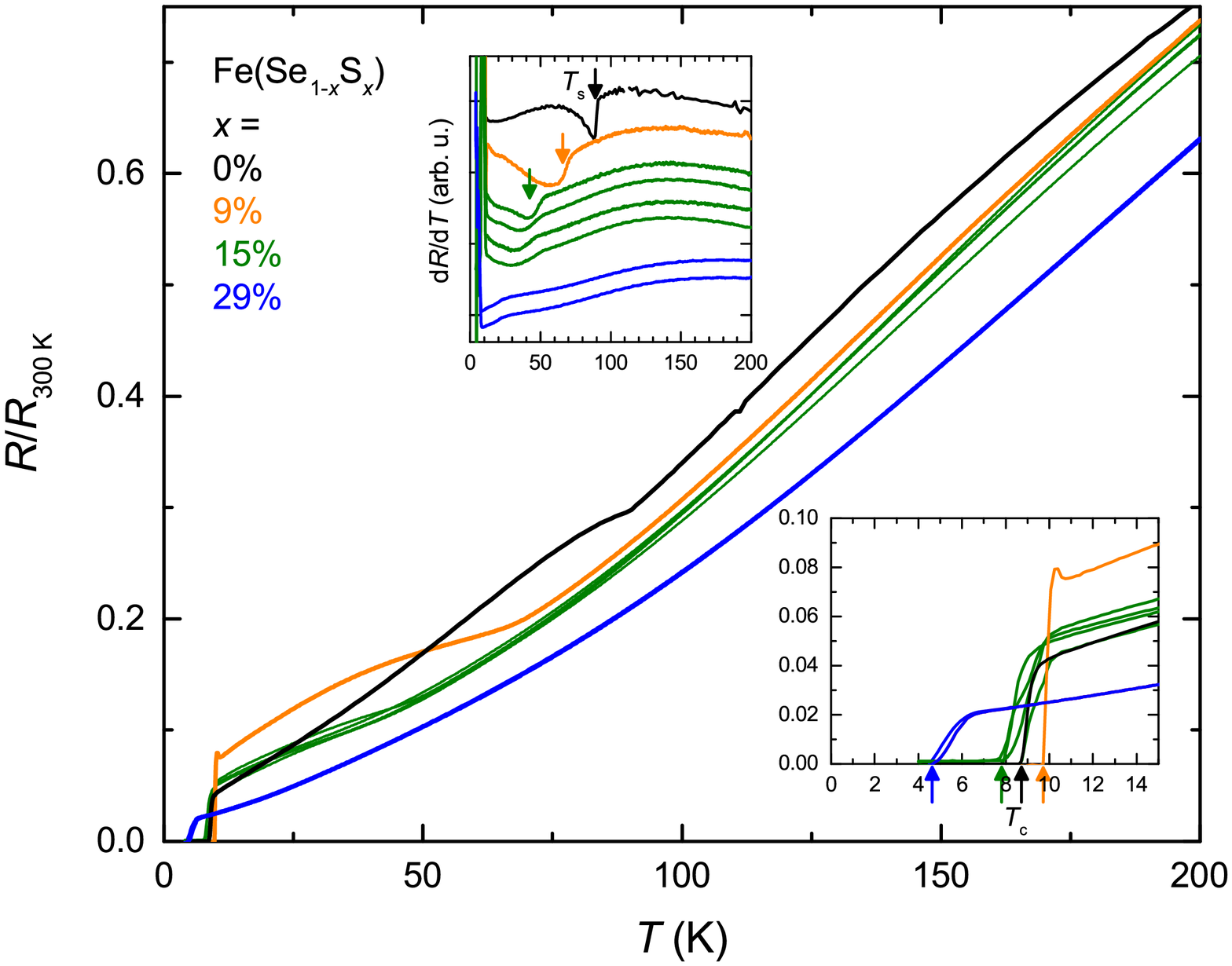}
	\caption{Electrical resistance of the investigated FeSe$_{1-x}$S$_x$ samples normalized at room temperature. A single sample with $x=0.09$, 4 samples with $x=0.15$ and a collection of $\sim30$ samples with $x=0.29$ were studied by NMR. For this highest sulfur content, the resistance of two representative samples is shown. For comparison, the resistance of an undoped FeSe single crystal \cite{Tanatar2016} is added. The upper inset shows the temperature derivative which is used to define $T_s$, the lower inset shows the low-temperature resistance on a magnified scale. $T_c$ is defined as the zero-resistance temperature.
	}
	\label{fig:RT}
\end{figure}\

\section{Sample growth and characterization}
The Fe(Se$_{1-x}$S$_x$) crystals were grown using chemical vapor transport similar to the description in Ref. \onlinecite{Bohmer2016}. Fe, Se and S powder were mixed in a ratio of 1.5:(1-$x_{\rm nom.}$):$x_{\rm nom}$ and sealed in a quartz ampoule together with a eutectic mix of AlCl$_3$ and KCl. The materials were let to react at 390$^\circ$C for 1-2 days before the ampoules were placed under a temperature gradient and chemical vapor transport was initiated. The sulfur content $x$, which varied from the nominal sulfur content $x_{\rm nom.}$, of several batches (shown as a full symbol in Fig. \ref{fig:latticeparameters}) was determined by energy-dispersive x-ray spectroscopy on 3-5 freshly cleaved crystals with an average of 4 different spots per crystal. The error bar indicates one standard deviation. Some early batches in which the initial reaction at 390$^\circ$C was omitted show quite substantial variations in sulfur content. The average lattice parameters of each batch were determined by powder x-ray diffraction in a Rigaku Miniflex diffractometer with Cu K$\alpha$ radiation. A minority hexagonal phase could sometimes be identified, however, the phase majority always was consistent with the tetragonal P4/nmm space group. The variation of lattice parameters with sulfur content is presented in Fig. \ref{fig:latticeparameters}. The results are consistent with findings in Ref. \onlinecite{Matsuura2017}. Furthermore, a linear extrapolation to $x=1$ yields almost perfectly the lattice parameters of FeS reported in Ref. \onlinecite{Lai2015}, indicating that Vegard's law is obeyed for the whole series. 

Figures \ref{fig:phasediag} and \ref{fig:RT} show the phase diagram of Fe(Se$_{1-x}$S$_x$) and the resistance data characterizing the NMR samples, respectively. The phase diagram reflects the variation in $x$ within some of the batches as individual samples of the same batch can exhibit varying $T_s$. To determine the sulfur content of the NMR samples as accurately as possible, we refer to the phase diagram, the EDS results and the average lattice parameters of the respective batches. The structural transition temperature $T_s$ of the samples was determined from resistivity measurements by the midpoint of the step in $dR/dT$ as in Refs. \onlinecite{Kaluarachchi2016,Xiang2017}. For the $x=0.09$ sample, EDS of a selection of samples from the same batch indicates $x=0.12(2)$. However, the resistance measurement in Fig. \ref{fig:RT} reveals that $T_s=68$ K for this sample. Thus, from the phase diagram, the specimen selected for NMR seems to be more accurately described by $x=0.09$. The four samples with $x\sim0.15$ show $T_s=41-48$ K in Fig. \ref{fig:RT}, located at $x=0.15$ in the phase diagram. The interpolation of lattice parameters for the batch yields $x=0.16$, in good agreement. For the samples with highest sulfur content, resistivity measurements do not find the signature of the structural phase transition and indicate $T_c\sim5$ K at zero field. The sulfur content $x=0.29$ is determined by interpolation of its lattice parameters, since the transition temperature $T_c$ barely varies with $x$ in this range and can therefore not be used as an indicator of sulfur content.

\section{Methods}
\subsection{NMR experimental details}
We conducted $^{77}$Se NMR ($I=1/2$; $\gamma/2\pi=8.118$ MHz/T) 
measurements under a fixed external field of $H=7.4089$ T applied either in the $ab$ plane or along the $c$ axis. 
The external field in the $ab$-plane was applied along the in-plane [110] tetragonal direction in order 
resolve the splitting of the NMR spectrum below $T_s$ for $H||ab$.
Measurements at $x=9\%$ were conducted on a single crystal of mass $\sim2$ mg, with $T_s\sim68$ K and $T_c\sim10$ K.
However, low NMR signal intensity prevented measurements above $\sim80$ K. 
To improve the signal intensity for the $x=15\%$ measurements, four single crystals of total mass $\sim10$ mg were each cleaved into 2 to 3 pieces. 
The samples were co-aligned by eye based on exterior faces of the crystals 
and  affixed to a glass plate with GE varnish. There was some variation of $T_s$ among this batch of crystals used for NMR measurements due to slight variations in sulfur content, with the average being $T_s=45\pm3$ K, see Figure \ref{fig:RT}. The variation is to be expected because the dependence of $T_s$ on doping is quite steep in this concentration range. All the samples showed $T_c\sim8$ K at zero field. 
For the $x=29\%$ measurements, $\sim30$ individual single crystals of total
mass $\sim35$ mg were fixed to a glass plate with GE varnish. The $ab$ plane orientation of the $x=29\%$ samples was not precisely controlled
as no nematicity was expected.

The $^{77}$Se NMR spin-lattice relaxation rate $1/T_1$ was measured with a recovery method using a single $\pi$/2 saturation pulse. The $1/T_1$ at each $T$ was determined by fitting the nuclear magnetization $m$ versus time $t$ using the exponential function
$1-m(t)/m(\infty)=\exp{(-t/T_1)}$,
 where $m(t)$ and $m(\infty)$ are the nuclear magnetization at time $t$ after the saturation and the equilibrium nuclear magnetization at $t \to \infty$, respectively.
 In the nematic state, no attempt was made to resolve the $T_1$ of the two peaks separately. We have measured only the $ab$ plane average $1/T_1$.
 NMR spectra were measured by FFT of the NMR spin echo. 
 
  \begin{figure}[t]
\centering
\includegraphics[width=\columnwidth]{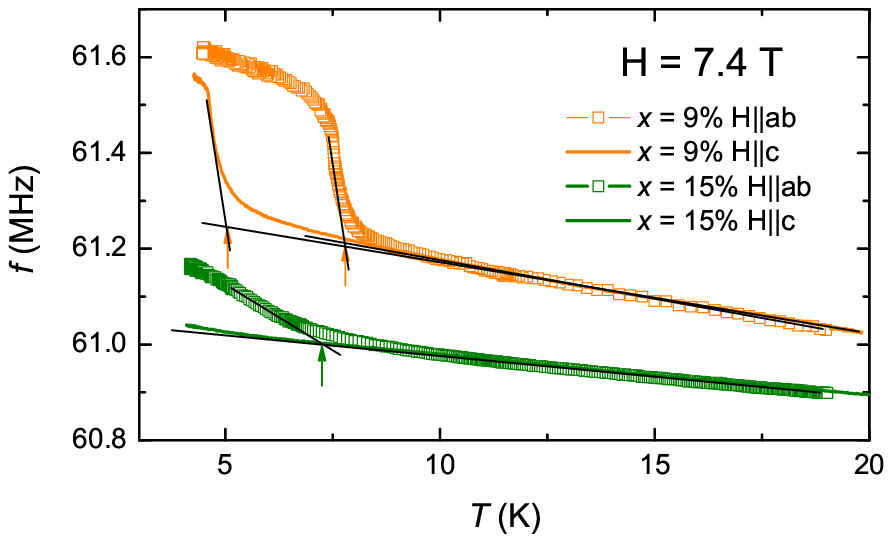}
\caption{{\it In situ} ac-susceptibility $\chi_{ac}$ measurement of $T_c$. 
The NMR coil tank circuit resonance frequency $f$ is a measure of $\chi_{ac}$ since $f=1/\sqrt{L_0(1+\chi_{ac})C}$.
Small arrows denote $T_c$ as determined by the intersection of two linear trends (black lines).
For $x=15\%$ $H||c$, $T_c$ is not observed above $T=4.0$ K. 
Data have been shifted vertically for clarity. 
Data for different samples have been obtained using different NMR coils, therefore the magnitude of the jump below $T_c$ cannot be directly compared. 
}
\label{fig:Tc}
\end{figure}
 
 The superconducting transition temperature
 $T_c(H)$ at the NMR measurement field ($H=7.4089$ T) was determined by {\it in situ} ac-susceptibility measurements down to $T=4.0$ K, as shown
 in Fig \ref{fig:Tc}. 
 The stronger suppression of $T_c$ for $H||c$ is consistent with Ref. \onlinecite{Abdel2015}.
 No superconductivity was observed above $T=4.0$ K at $H=7.4089$ T for $x=29\%$ samples.
 
 \subsection{Scaling analysis of $1/T_1T$ and $K_{\rm spin}^2$}

In the main paper we decomposed $1/T_1T=(1/T_1T)_{\rm AFM}+(1/T_1T)_{\mathbf{q}=0}$.
The $\mathbf{q}=0$ term will show Korringa behavior 
\begin{equation}
(1/T_1T)_{\mathbf{q}=0}=CK_{\rm spin}^2.
\label{eq:prop}
\end{equation}
To obtain $K_{\rm spin}=K-K_0$ one needs the chemical shift $K_0$, which is obtained from a $K$ vs. $\chi$ plot analysis (see below).
The proportionality constant in Eq. \ref{eq:prop} is given by $C=\alpha S^{-1}$,
where $S$ is the Korringa constant $S=(\hbar/4\pi k_{\rm B})\left(\gamma_{\rm e}/\gamma_{\rm N}\right)^2$ ($S=7.23\times10^{-6}$ Ks for $^{77}$Se)
and the Korringa ratio $\alpha$ parameterizes deviations of $C$ from the theoretical value $S^{-1}$ \cite{Wiecki2015}.

The necessary scaling factors $C$ were empirically determined from a plot of $(1/T_1T)_i$ against $K_{{\rm spin},i}^2$ ($i=\perp,\|$) with $T$ as an implicit parameter.
The points above $T=90$ K showed linear behavior, the slope of which determines $C$.
Since we measure $K_{{\rm spin}}$ in units of $\%$, this analysis determines determines $C$ in units of ($\%)^{-2}(Ks)^{-1}=10^4(Ks)^{-1}$,
as reported in the main paper.
For the $x=9\%$ sample, we assumed the same values of $C$ and $K_0$ as for $x=15\%$ since
we lack the high-$T$ data due to low signal intensity.

\section{Additional data}

\subsection{$(1/T_1T)_{AFM}$ Contour Plot}

\begin{figure}[t]
\centering
\includegraphics[width=\columnwidth]{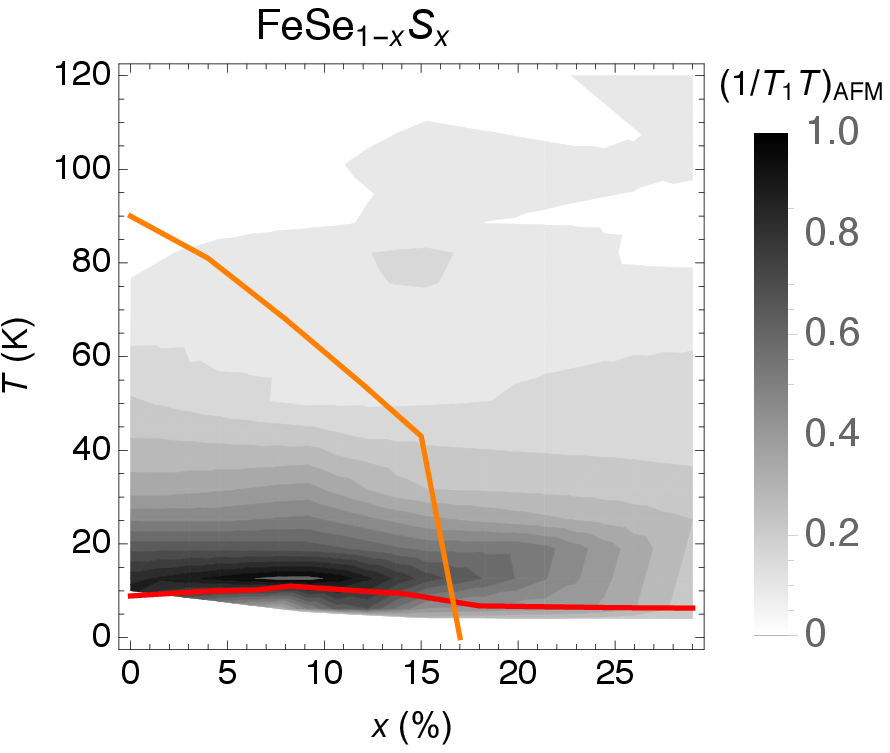}
\caption{Contour plot of AFM fluctuations in FeSe$_{1-x}$S$_x$. Here we define $(1/T_1T)_{\rm AFM}\equiv(1/T_1T)_{\|}-CK_{{\rm spin},c}^2$
in contrast to the main paper. 
}
\label{fig:contour}
\end{figure}

\begin{figure}[b]
\centering
\includegraphics[width=\columnwidth]{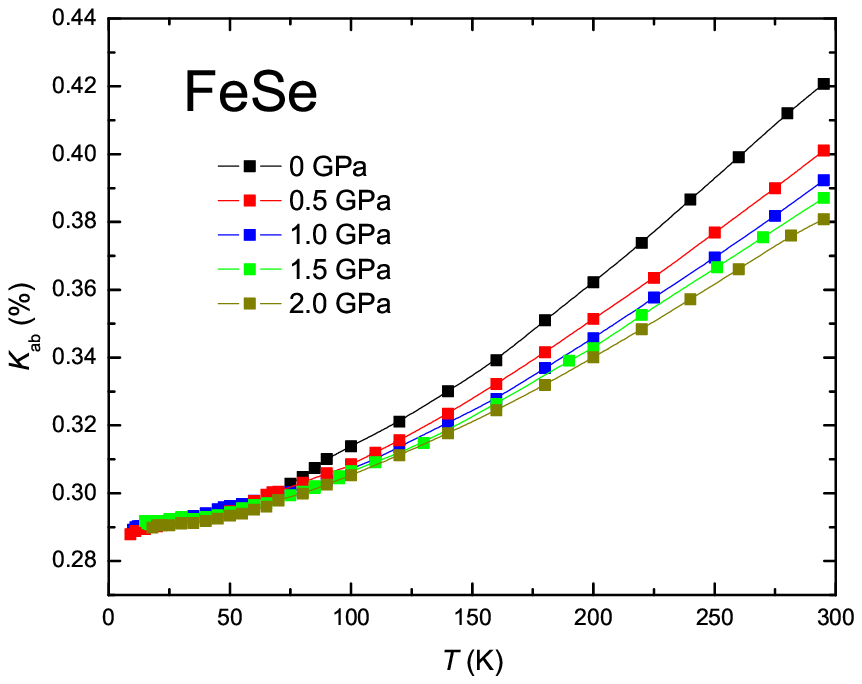}
\caption{NMR shift with $H||ab$ in pure FeSe under pressure.}
\label{fig:KabFeSe}
\end{figure}

In the final plot of the main paper, we compared the AFM contribution to $1/T_1T$ in pressurized and sulfur-doped FeSe.
Obviously, we would like to compare the same quantity for both systems. 
However, in the case of pressurized FeSe, we lack data for $H\|c$ making the full analysis involving $1/T_{1,\|}T$ and $1/T_{1,\perp}T$ impossible.
We therefore simply used the definition $(1/T_1T)_{\rm AFM}\equiv(1/T_1T)_{H\|ab}-CK_{{\rm spin},ab}^2$ for the comparison.
However, as we describe in the paper, it is preferable to define $(1/T_1T)_{\rm AFM}\equiv(1/T_1T)_{\|}-CK_{{\rm spin},c}^2$, 
which consistently compares the susceptibility of the $c$-axis component of the hyperfine field. We include this contour plot here for comparison in Fig \ref{fig:contour}. 
The plot is qualitatively similar to the one used in the main text and thus the definition of $(1/T_1T)_{\rm AFM}$ has no effect on our physical conclusions. 

\subsection{NMR Shift in FeSe under pressure}

Fig. \ref{fig:KabFeSe} shows the NMR shift with $H||ab$ in pure FeSe under pressure. 
Here, for simplicity, we show the $ab$-plane average $K_{ab}=(K_a+K_b)/2$ in the orthorhombic phase.
As in the case of S doping in the main paper, $K_{ab}$ is pressure independent at low $T$, but depends on pressure at high $T$. 
Here, the high-$T$ value of $K_{ab}$ decreases with increasing pressure, similar to the behavior with increasing S doping.

\subsection{$K$ vs $\chi$ Analysis}

\begin{figure}[t]
\centering
\includegraphics[width=\columnwidth]{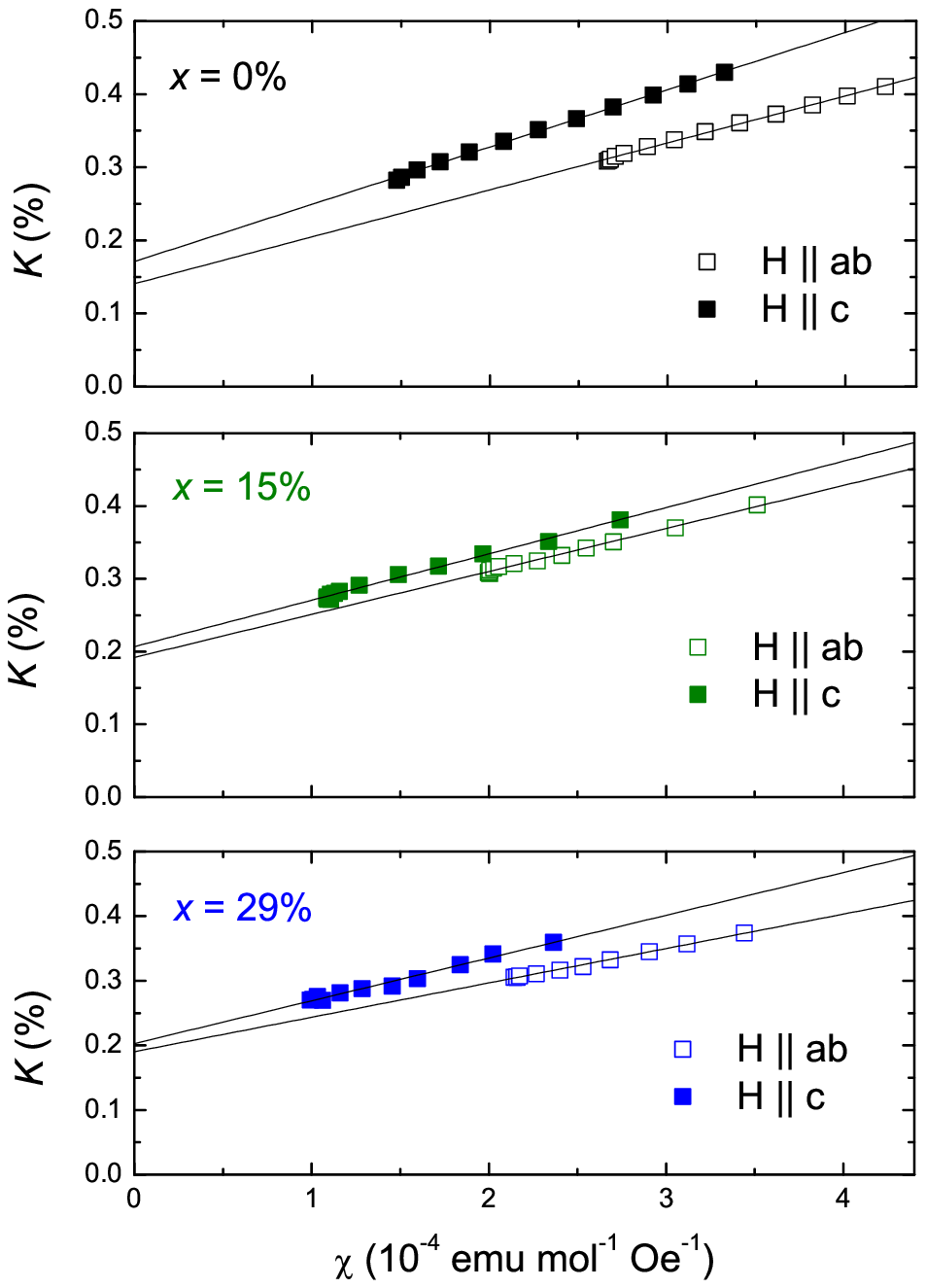}
\caption{$K$ vs $\chi$ plot analysis for $x=0$, $x=15\%$ and $x=29\%$. 
}
\label{fig:kchi}
\end{figure}

We performed a $K$ vs $\chi$ plot analysis to determine the $T$-independent chemical shift $K_0$ and 
hyperfine coupling constants $A_{ab}$ and $A_c$. In Fig. \ref{fig:kchi}, we plot $K$ as a function of $\chi$ with $T$ as an implicit parameter. The $T$ range is chosen so as to avoid 
low-$T$ upturns of $\chi$ due to magnetic impurities to which NMR, a local probe, is insensitive.
For $x=9\%$, such an analysis is not possible because we lack data over the entire $T$ range due to signal 
intensity problems at high $T$. In these plots, $K_0$ is the $y$-intercept. The hyperfine coupling constants
are determined by the slope. 
For $x=0\%$, we obtain $A_{ab}=3.585$ T/$\mu_B$ and $A_{c}=4.37$ T/$\mu_B$. 
For $x=15\%$, we obtain $A_{ab}=3.3$ T/$\mu_B$ and $A_{c}=3.56$ T/$\mu_B$. 
For $x=29\%$, we obtain $A_{ab}=2.97$ T/$\mu_B$ and $A_{c}=3.7$ T/$\mu_B$.


\begin{thebibliography}{99}
\bibitem{T1_ab}   Here, $(1/T_1T)_{H||ab}$ is averaged over any $ab$ plane anisotropy due to nematicity \cite{SM}.
\bibitem{Yongbin}
   For obtaining self-consistent charge density, we employed muffin-tin radii $R_{\rm MT}=2.0$ a.u. for Fe and Se, respectively, and  $R_{\rm MT}k_{\rm max}=8.0$ and  which set the maximum $k$ value in the expansion of plane waves. 
The Se atom position was relaxed until the force on the Se atom was less than 0.1 mRy/a.u.
Calculations were iterated, with 4800 $k$-points for whole Brillouin zone, to reach the total energy convergence criterion that was 0.01 mRy/cell.
For the Fermi surface calculations, we choose we chose $k_{z}=0.0 (2\pi/c)$  plane and divide  $-2\pi/a \leq k_{x}, k_{x} \leq 2\pi/a$ region by 200*220 mesh that resulted in 40401 k-points.
\end{thebibliography}
\end{document}